# Interference at the tunnel ionization from double-center potential

P.A. Golovinski[1,2], A.A. Drobyshev[2]

[1] *Moscow Institute of Physics and Technology, Moscow 141700, Russia*
[2] *Physics Research Laboratory, Voronezh State University of Architecture and Civil Engineering, Voronezh 394006, Russia*

E-mail: golovinski@bk.ru

The tunnel ionization of an electron bounded by two delta potentials under the influence of a constant electric field is considered. The equations for the electron current density for two different initial states are obtained. The dependence of the emission current on the orientation of the potential with respect to the field direction and the distance between them is studied. Appropriate conditions for observation essential interference effects are defined.

### 1. Introduction

The model of a short-range potential is widely used in atomic physics, nuclear physics, and in condensed matter physics [1-3]. For the problem of the tunnel ionization of neutral molecules the generalization of asymptotic method [4] for the wave functions in parabolic coordinates in spherically asymmetric states has been developed [5-7]. However, in this case the theory is limited to consideration of the localized states of individual molecular orbitals. The concept of localized states without taken into account possible interference lies at the basis of the description of tunneling effects in nanostructures [8, 9]. At the same time, detecting the possible role of delocalization initial states requires special consideration.

The problem of the particle dynamics in the field of several short-range potentials allows implementation of a complete analytical study [2, 10]. When there is a close located potential centers, the wave functions overlap, and formed a common state. In addition to changing the structure the stationary states itself, such systems have a specific response to the external field [11]. We have considered the interference pattern for electron waves generated by electron tunneling ionization from double-center potential with a different field and potential orientation.

## 2. Basic equations

The problem of the particle dynamics in a field a few $\delta$-potentials can be formulated in terms of the boundary conditions imposed on the wave function in the potential location points. For a single $s$-state with $l=0$, the wave function $\psi \sim r^{-1} \exp(-\alpha r)$, and the boundary condition takes the form [1]

$$\left. \frac{d \ln(r\psi)}{dr} \right|_{r=0} = -\alpha, \qquad (1)$$

where $\alpha = \sqrt{-2E_0}$, $E_0$ is bound energy. We use the atomic system of units in which $|e| = m = \hbar = 1$.

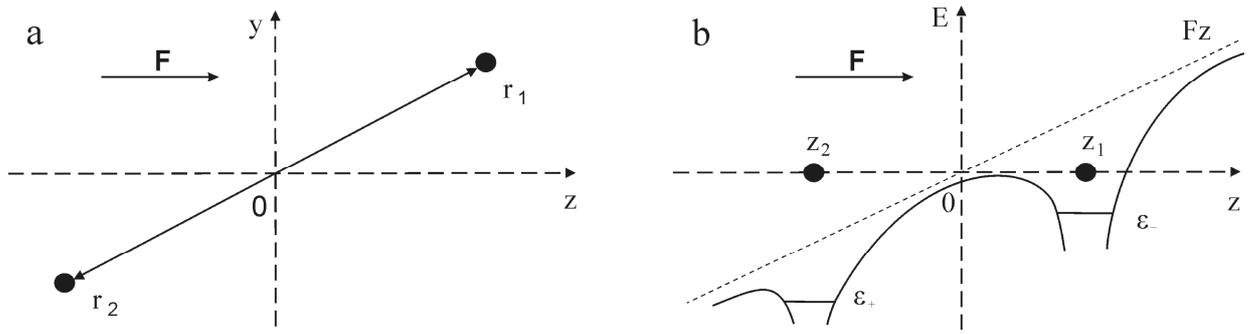

**Fig. 1.** The relative location of the center (a) and the position of the energy levels for the system orientation parallel to the field (b).

For a few potentials their action is provided by the sum of individual potentials. State of a particle in the field of two identical attraction centers and an external electric field of strength $F > 0$ (Fig. 1) is described by the stationary Schrödinger equation

$$\left(\varepsilon + \frac{1}{2}\nabla^2 + Fz\right)\psi(\mathbf{r}) = \left(V_1(\mathbf{r}-\mathbf{r}_1) + V_2(\mathbf{r}-\mathbf{r}_2)\right)\psi(\mathbf{r}), \qquad (2)$$

where each potential [3]

$$V(\mathbf{r}) = -\frac{2\pi}{\alpha}\delta(\mathbf{r})\frac{\partial}{\partial r}r. \qquad (3)$$

The solution, as in the problem without the field [12], is sought in the form of a superposition of the Green's functions

$$\psi(\mathbf{r}) = AG(\mathbf{r},\mathbf{r}_1,\varepsilon) + BG(\mathbf{r},\mathbf{r}_2,\varepsilon), \quad (4)$$

where $G(\mathbf{r},\mathbf{r}_j,\varepsilon)$ is the Green's function with the asymptotic behavior of the outgoing waves, satisfying the Eq. (2) with the replacement of the right side to $\delta(\mathbf{r}-\mathbf{r}_j)$, $\mathbf{r}_1$ is the radius vector of the first center, $\mathbf{r}_2$ is the radius vector of the second center. The Green's function, for a particle moving under the action of a constant force [13], is expressed in terms the Airy function [14]:

$$G(\mathbf{r},\mathbf{r}',\varepsilon) = \frac{1}{2|\mathbf{r}-\mathbf{r}'|}\{Ci(\chi_+)Ai'(\chi_-) - Ci'(\chi_+)Ai(\chi_-)\}, \quad (5)$$

where

$$\chi_\pm = -\frac{1}{(2F)^{2/3}}\left(2\varepsilon + \mathbf{F}(\mathbf{r}+\mathbf{r}') \pm F|\mathbf{r}-\mathbf{r}'|\right), \quad (6)$$

and $Ci(u) = Bi(u) + iAi(u)$ has the asymptotic behavior of the outgoing wave at $u \to \infty$.

Taking into account the boundary conditions of the form of Eq. (1), we obtain a system of equations for the complex quasi-energy $\varepsilon = \operatorname{Re}\varepsilon - i\Gamma/2$ and the coefficients $A$ and $B$:

$$\frac{1}{\rho_1\psi}\frac{\partial}{\partial\rho_1}(\rho_1\psi)_{\rho_1=0} = -\alpha,$$
$$\frac{1}{\rho_2\psi}\frac{\partial}{\partial\rho_2}(\rho_2\psi)_{\rho_2=0} = -\alpha. \quad (7)$$

Here we use the notation $\rho_1 = \mathbf{r}-\mathbf{r}_1$, $\rho_2 = \mathbf{r}-\mathbf{r}_2$.

The Eqs. (7) can be written as [15]

$$\begin{pmatrix} b_+ & G(-\mathbf{R}/2,\mathbf{R}/2,\varepsilon) \\ G(\mathbf{R}/2,-\mathbf{R}/2,\varepsilon) & b_- \end{pmatrix}\begin{pmatrix} A \\ B \end{pmatrix} = 0, \quad (8)$$

$$b_\pm = \frac{\pi}{(2F)^{1/3}}\left(Ai'(\xi_\pm)Ci'(\xi_\pm) - \xi_\pm Ai(\xi_\pm)Ci(\xi_\pm)\right) + \alpha, \quad (9)$$

$$\xi_\pm = -\frac{2\varepsilon \pm FR\cos\theta}{(2F)^{2/3}},$$

where $\mathbf{r}_1 = \mathbf{R}/2$, $\mathbf{r}_2 = -\mathbf{R}/2$, $R = |\mathbf{r}_1 - \mathbf{r}_2|$. From Eq. (8) follows the coupling of coefficients $A$ and $B$:

$$\frac{A}{B} = -\frac{2\pi G(-\mathbf{R}/2, \mathbf{R}/2, \varepsilon) + b_\pm}{2\pi G(\mathbf{R}/2, -\mathbf{R}/2, \varepsilon) + b_\mp}\bigg|_{\varepsilon = \varepsilon_\pm}. \qquad (10)$$

Eq. (7) has two solutions, defining two wave function and two energy values [15]

$$\varepsilon_\pm = E_\pm + \Delta\varepsilon_\pm - \frac{i}{2}\Gamma_\pm \qquad (11)$$

respective states with their widths, that are found from the condition that the determinant of the system (8) is equal zero. Here, $E_\pm = -k_\pm^2$ are unperturbed energies of the symmetric and antisymmetric states of the system [12] and

$$\kappa_\pm = \alpha \pm \frac{e^{-\kappa_\pm R}}{R}. \qquad (12)$$

For a weak electric field ($F/\alpha^3 \ll 1$) the level shifts and width were calculated earlier [11].

### 3. Numerical simulations

The observed value in the process of ionization is electron current density crossing the plane perpendicular to the z-axis, defined by the direction of the electric field:

$$j_z = \mathrm{Im}\left(\psi^* \frac{\partial \psi}{\partial z}\right). \qquad (13)$$

Spatial electron current distribution depends on the angle between the axis connecting the centers, and the direction of the electric field, parameter $\alpha|\mathbf{r}_1 - \mathbf{r}_2|$ characterizing the distance between the centers in comparison with attenuation unperturbed wave function localized at a single center, and parameter $F/\alpha^3$ that defines the magnitude of the force action of the electric field with respect to the characteristic force that bounds the particle in the potential well.

There are two limiting cases of double-center potential orientation with respect to the field: the field parallel and perpendicular to the field. At the orientation centers of attraction perpendicular to the field ($\theta = \pi/2$), the ratio of the coefficients $|A/B|=1$. This means that the electron wave function is a superposition of the wave functions each $\delta$-potential and an electron is localized on equally both centers. Increasing the length and the field strength leads to decrease in the displacement of states, and the electron is localized at one of the centers. In the ground state electron is localized on the second center and tunnels without influence of the first center, and a tunnel current distribution is analogous to the current from a single point source.

Fig. 2 shows the result of calculating the energy of the ground (excited) state and the ratio of the coefficients $|A/B|$ depending on the angle $\theta$ between **R** and **F** for the parameters $E_0 = 1\,\text{eV}$, $F = 10^7\,\text{V/cm}$, $R = 10$ a.u.

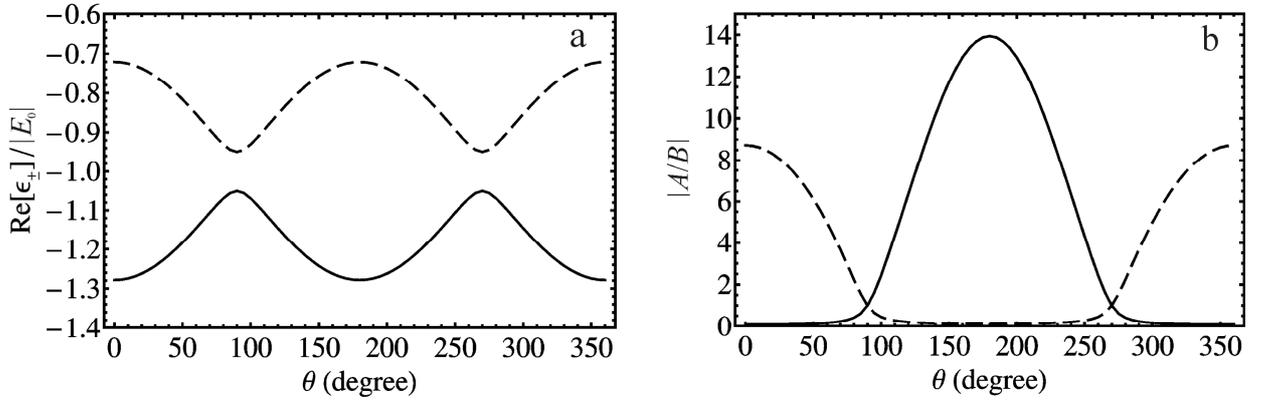

**Fig. 2.** Energy dependence $\varepsilon_\pm$ (a) and the ratio dependence $|A/B|$ of the coefficients (b) on the angle. Solid line is the ground state, dotted is excited state.

From the results plotted in the figure it is clear, that for $\theta = \pi/2$ the ratio of the coefficients $|A/B|=1$ in the ground and excited states, which corresponds to the symmetry. When the orientation of the system is parallel to the field ($\theta = 0$), in the ground state $|A/B|=0.07$, i.e., electron actually is localized on the second center. In the excited state, when $\theta = 0$, the ratio of the coefficients $|A/B|=9$, corresponding to the electron localization at the first center.

### 3. Conclusions

The considered interference of electron waves generated under the electron tunnel ionization from the double-center potential essentially depends on the orientation of the field and potential due to the degree of electron localization near the different centers. When the orientation of the double-center potential is perpendicular to the field, the distribution of the current is equivalent to the electron current distribution pattern for two coherent point sources. When one have the inclined orientation of the system, the delocalization of the initial states increases with increasing the field strength and the distance between the centers. The calculations indicate the importance of taking into account the type of bonds in the interpretation of the results of observation of the electron tunnel ionization.

This work is supported by RFBR (grant №16-32-00253)